\documentclass[universe,article,accept,pdftex,moreauthors]{Definitions/mdpi} 

\firstpage{1} 
\pubvolume{1}
\issuenum{1}
\articlenumber{0}
\pubyear{2023}
\copyrightyear{2023}
\externaleditor{Academic Editor: Name}
\datereceived{2 June 2023} 
\daterevised{12 July 2023} 
\dateaccepted{13 July 2023} 
\datepublished{ } 
\hreflink{https://doi.org/} 

\usepackage{natbib}

\Title{Estimates of the Surface Magnetic Field Strength of \mbox{Radio Pulsars}}

\TitleCitation{Estimates of the Surface Magnetic Field Strength of Radio Pulsars}

\Author{Vitaliy Kim 
 $^{1,2,}$*\orcidA{}, Adel Umirbayeva $^{1,3,\dagger}$\orcidB{} and Yerlan Aimuratov $^{1,3,4,}$*$^{,\ddagger}$\orcidC{}}

\AuthorNames{Vitaliy Kim, Adel Umirbayeva and Yerlan Aimuratov}

\AuthorCitation{Kim, V.; Umirbayeva, A.; Aimuratov, Y.}

\address{%
$^{1}$ \quad Fesenkov Astrophysical Institute, Observatory 23, Almaty 050020, Kazakhstan\\
$^{2}$ \quad Pulkovo Observatory, Pulkovoskoe Shosse 65/1, 196140 Saint Petersburg, Russia\\
$^{3}$ \quad Dept. of Theoretical and Nuclear Physics, Faculty of Physics and Technology, Al-Farabi Kazakh National University, Al-Farabi Avenue 71, Almaty 050040, Kazakhstan\\
$^{4}$ \quad International Center for Relativistic Astrophysics Network, Piazza della Repubblica 10, 65122 Pescara, Italy}

\corres{Correspondence: kim@fai.kz (V.K.); aimuratov@fai.kz (Y.A.)}

\firstnote{Currently a Master student at Al-Farabi KazNU.} 
\secondnote{Currently a Postdoctoral associate at Al-Farabi KazNU.}

\abstract{ We investigate the geometry of the magnetic field of rotation-powered pulsars. A new method for calculating an angle ($\beta$) between the spin and magnetic dipole axes of a neutron star (NS) in the ejector stage is considered within the frame of the magnetic dipole energy loss mechanism. We estimate the surface magnetic field strength ($B_{\rm ns}$) for a population of known neutron stars in the radio pulsar (ejector) stage. The evaluated $B_{\rm ns}(\beta)$ may differ by an order of magnitude from the values without considering the angle $\beta$. It is shown that $B_{\rm ns}(\beta)$ lies in the range $10^{8}$--$10^{14}\,\text{G}$ for a known population of short and middle periodic radio pulsars.
}


\begin{document}
\section{Introduction}
\label{sec:introduction}

\begin{figure}[H]
\includegraphics[width=7.5cm]{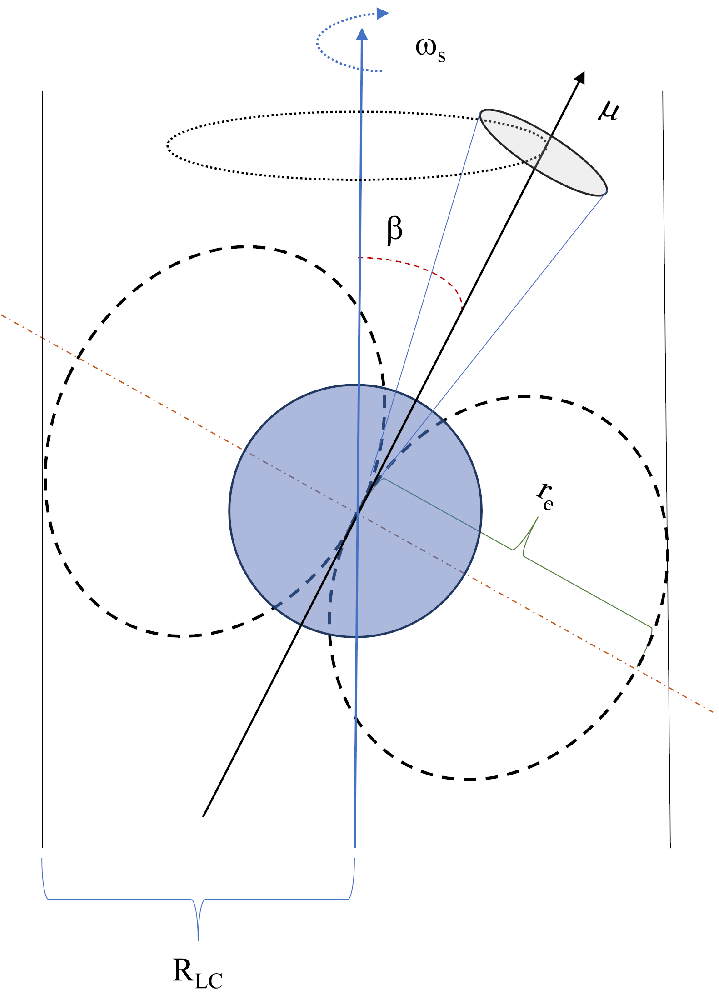}
\caption{Scheme of rotating magnetized neutron star and its axes. In the figure denoted: $\mu$ is a magnetic dipole moment of NS, $\omega_{\rm s}$ is an angular rotation velocity, $\beta$ is an angle between spin and magnetic dipole axes, $r_{\rm e}$ is an equatorial radius of the magnetic field of NS, and $R_{\rm LC}$ is a radius of the light cylinder of NS.}
\label{fig2}
\end{figure}

Radio pulsars are fast-spinning magnetized neutron stars (NS) demonstrating regular modulations (pulsations) of their radiation with a high stable period in the radio range. The axis of the magnetic field of the radio pulsar and its spin axis are not aligned, and the beam of radiation is emitted in a cone-shaped region (see Figure~\ref{fig2}). Therefore, pulsar radiation is seen as pulses (beacon effect) by an external observer \cite{Lyne...2012puas.book.....L}. 

Radio pulsars are characterized by the rapid axial rotation (or spin) they have acquired due to the conservation of angular momentum during their formation. Their spin periods ($P_{\rm s}$) lie in a wide range: from $0.0014$~s to $23.5$~s \cite{Manchester...2005AJ...129...1993M, Tan...2018ApJ...866...54T}, with the majority not exceeding a few seconds. Radio pulsars are usually divided into several groups depending on their spin period \cite{Malov...2022ARep...66...25M, Kenko...2023MNRAS.tmp.1064K}:

\begin{enumerate}
    \item Short-periodic pulsars, including millisecond pulsars ($P_{\rm s} < 0.1 \text{s}$);
    \item Middle-periodic pulsars ($0.1 \text{s} < P_{\rm s} < 2 \text{s}$);
    \item Long-periodic pulsars ($ P_{\rm s} > 2 \text{s}$).
\end{enumerate}

According to \cite{Kenko...2023MNRAS.tmp.1064K}, a pulsar wind mainly causes the spin-down process of long-periodic radio pulsars. However, for short and middle periodic radio pulsars, a primary mechanism of rotation energy loss ($\dot{E}_{\rm obs}$) is believed to be magnetic dipole radiation (MDR).

The MDR mechanism of the energy loss was first considered in \cite{Gunn...1969Natur.221..454G, Pacini...1970Natur.226..622P} for radio pulsars. It was shown that the magnetized NS could lose its rotational energy by MDR generation. This evolution stage of NS is also known as the ``ejector'' stage, and its energy loss ($\dot{E}_{\rm md}$) for the generation of MDR expresses as:

\begin{linenomath}
\begin{equation}
\dot{E}_{\rm md} = -\frac{2}{3}\,\frac{\mu_{\rm ns}^2\, \omega^{4}_{\rm s}\,(\sin{\beta})^2}{c^3}~,
\label{eq:loss}
\end{equation}
\end{linenomath}
where $\mu_{\rm ns} = B_{\rm ns}\,R_{\rm ns}^3/2$ is the magnetic dipole moment of NS, $R_{\rm ns}$ is the radius of NS, $\omega_{\rm s} = 2\pi/P_{\rm s}$ is the angular rotation velocity, $c$ is the speed of light, and $\beta$ is the angle between spin and magnetic dipole axes; the value of $\beta$ lies within $0$--$90$~deg.

The expression for rotational energy loss is the following:

\begin{linenomath}
\begin{equation}
\dot{E}_{\rm obs} = I\omega_{\rm s} \dot{\omega}_{\rm s} = -I\frac{4 \pi^2 \dot{P}_{\rm s}}{P_{\rm s}^3}~,
\label{eq:lossrotation}
\end{equation}
\end{linenomath}
where $I$ is a moment of inertia of NS, $\dot{\omega}_{\rm s} = - 2\pi \dot{P}_{\rm s}/P_{\rm s}^2$ is a derivative of the angular rotation velocity, $\dot{P}_{\rm s}$ is a derivative of the spin period, i.e., rotational spin-up or spin-down.

Solving the system of Equations~(\ref{eq:loss})--(\ref{eq:lossrotation}) by equating the losses, with $\mu_{\rm ns}$ and using canonical values for NS (see Section~\ref{subsec:materils}), one can derive an expression for $B_{\rm ns}$ as

\begin{linenomath}
\begin{equation}
B_{\rm ns} \sin{\beta} = 3.2 \times 10^{19} \sqrt{P_{\rm s} \dot{P}_{\rm s}}~~~~{\rm G}~.
\label{eq:magneticbeta}
\end{equation}
\end{linenomath}

Equation~(\ref{eq:magneticbeta}) is a basic expression for estimating the magnetic field strength for rotation-powered pulsars. As seen from the equations above, $\dot{E}_{\rm md}$ significantly depends on angle $\beta$, reaching the maximal energy loss $\dot{E}_{\rm md}^{\rm (max)}$ with orthogonal axes ($\beta = 90^{\circ}$). Indeed, in most cases, the magnetic field strength for radio pulsars is estimated by accepting the $\dot{E}_{\rm md}^{\rm (max)}$ case; however, angle $\beta$ can vary widely from the maximum value. Thus, it is crucial to correctly estimate the angle between spin and magnetic dipole axes to evaluate $B_{\rm ns}(\beta)$ for rotation-powered pulsars.

\vspace{-6pt}

Various methods for estimating the $\beta$-parameter have been previously proposed in the literature \cite{Kuzmin...1983PAZh....9..149K, Lyne...1988MNRAS.234..477L, Malov...1990SvA....34..189M, Kenko...2022ARep...66..669K}. Here, we offer a relatively simple method based on a geometric approach for calculating the angle between the spin and magnetic dipole axes of a neutron star (NS) in the ejector stage. Section~\ref{subsec:meth} outlines a basic concept and geometry for $\beta$. On its basis, in Section~\ref{subsec:materils}, we evaluate the surface magnetic field strength $B_{\rm ns}(\beta)$ for a population of known neutron stars from the ATNF pulsar catalog. We provide the main results with a discussion in Sections~\ref{sec:results} and~\ref{sec:discussion}.

\textls[-15]{The obtained data can help study properties and geometry of NS magnetic fields \cite{Igoshev...2021Univ....7..351I}, study and model pulsar spin evolution \cite{Xie...2014AN....335..775X}, investigate stellar evolution in the late stages \cite{Portegies...1999NewA....4..355P}, etc.}

\section{Methods}
\label{sec:mat_meth}

\subsection{Estimation of $\beta$-Parameter}
\label{subsec:meth}

\begin{figure}[H]
\includegraphics[width=9cm]{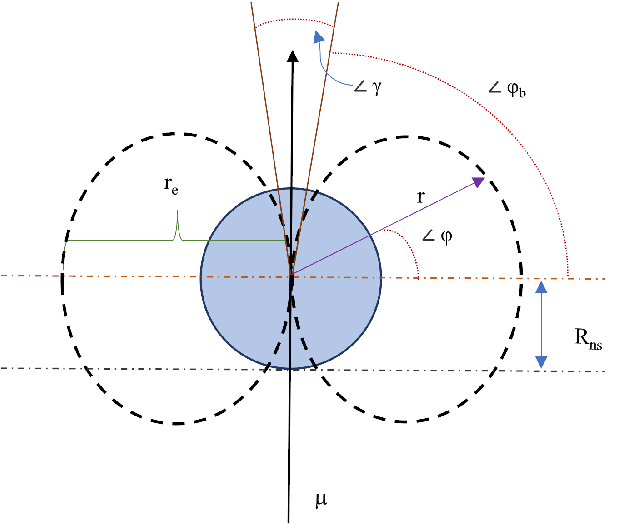}
\caption{Scheme of a neutron star magnetosphere. In the figure denoted: $R_{\rm ns}$ is a radius of the neutron star, $r$ is a radius vector of the magnetosphere, $\phi$ is an angle measured from the magnetic equator $r_{\rm e}$ towards the magnetic pole, with $\mu$ being an axis of a magnetic dipole moment of NS, $\gamma$ is an opening angle of emission cone, $\phi_{\rm _b}$ corresponds to the angle between the magnetic equator and lateral surface of the emission cone, and $r_{\rm e}$ is an equatorial radius of the magnetic field of NS.}
\label{fig1}
\end{figure}

According to \cite{Lipunov...1992ans..book.....L}, an equation of magnetic field lines, based on an assumption of a dipole magnetic field, in polar coordinates expressed as
\begin{equation}
r = r_{\rm e}\, (\cos{\phi})^2~,
\label{eq:polarforcelinemf}
\end{equation}
where $r_{\rm e}$ is an equatorial radius of the magnetic field, corresponding to NS magnetosphere radius (see Figure~\ref{fig1}), $\phi$ is an angle measured from the magnetic equator $r_{\rm e}$ towards the magnetic pole. 

\vspace{-6pt}

Using an Equation~(\ref{eq:polarforcelinemf}), it is possible to find an opening angle of emission cone $\gamma$ assuming that the radio pulsar magnetosphere is limited by a light cylinder with radius $R_{\rm LC}$, i.e., $r_{\rm e} = R_{\rm LC} = c \times P_{\rm s} / 2\pi$ \cite{Goldreich...1969ApJ...157..869G}. At the base of the emission cone, the radius vector of the magnetic field line corresponds to the NS radius $r_{\rm b} = R_{\rm ns}$. 

Solving Equation~(\ref{eq:polarforcelinemf}), we can find $\phi_{\rm _b}$ corresponding to the angle between the magnetic equator and lateral surface of the emission cone (see Figure~\ref{fig1}).
\begin{equation}
\phi_{\rm _b} = \arccos{\left[\left( \dfrac{R_{\rm ns}\, \times 2\pi}{c\times P_{\rm s}}\right)^{1/2}\,\right]}~.
\label{eq:phisolution}
\end{equation}

Using Equation~(\ref{eq:phisolution}), one can find an opening angle of the emission cone $\gamma$: 
\begin{equation}
\gamma = 2\times(90^{\circ} - \phi_{\rm _b})~.
\label{eq:gammarange}
\end{equation}

\begin{figure}[H]
\includegraphics[width=0.9\textwidth]{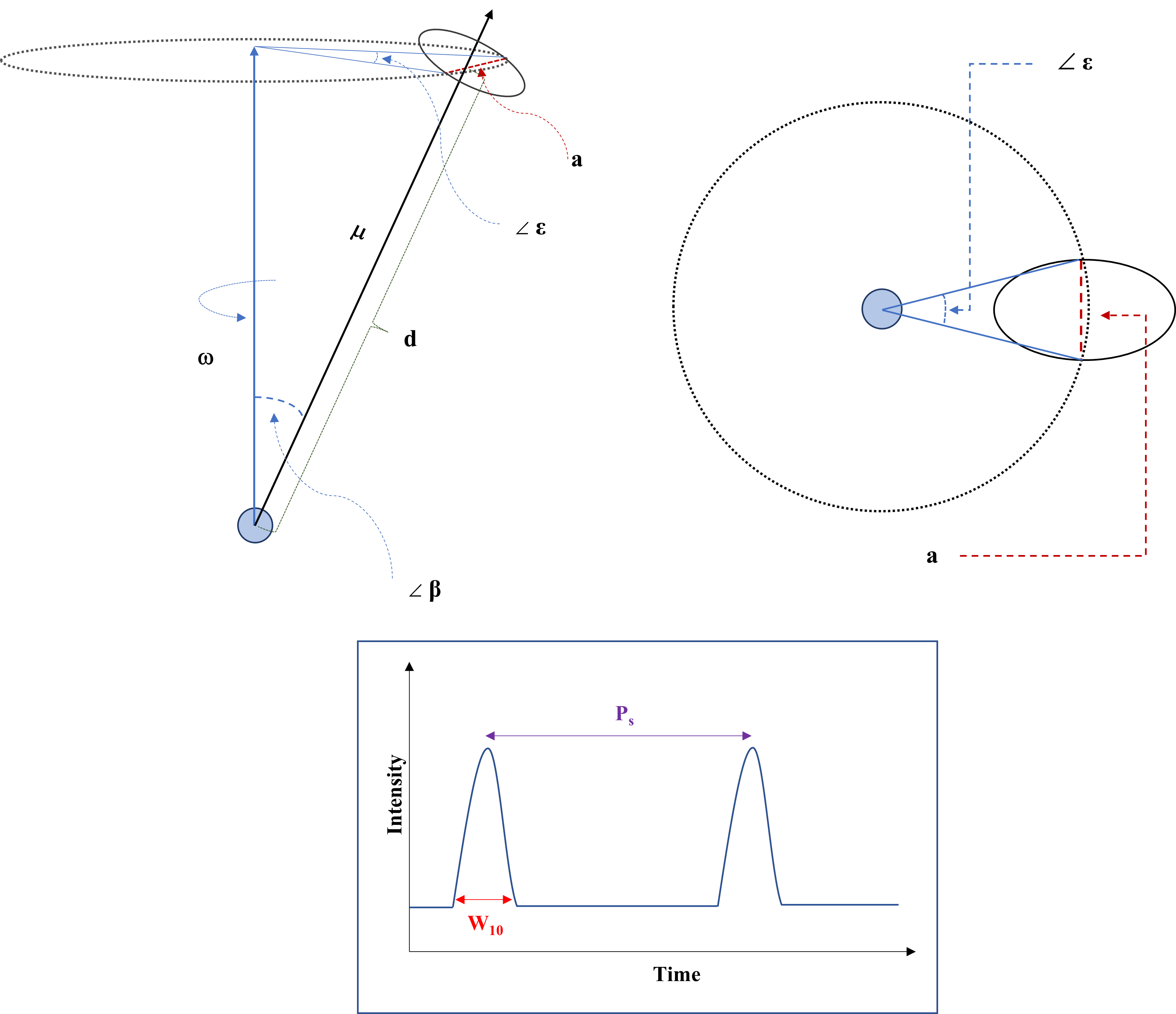}
\caption{Scheme of the opening angle of the emission cone and the angle $\beta$ between spin and magnetic dipole axes (\textbf{left} panel), and its view from the top (\textbf{right} panel). In the figure denoted: $a$ is the linear diameter of the emission cone, and $d$ is the distance.  The rest of the symbols are identical to the ones in Figures~\ref{fig2} and \ref{fig1}. Bottom panel shows the pulse profile with period ($P_{\rm s}$) and width of individual pulse at $10\%$ of maximal intensity ($w_{10}$).}
\label{fig3}
\end{figure}

At the next step, one can consider the dihedral angle (Figure~\ref{fig3}) formed by two emission cone guides $d$ and the rotation axis $\omega_{\rm s}$. Then, the linear diameter $a$ of the emission cone at a distance $d$ will be expressed as follows (via the cosine theorem):

\begin{equation}
\begin{cases}
a^2 =   2d^2 (1 - \cos{\gamma})\\[0.1cm]
a^2 =  2(d \sin{\beta})^2 \,(1 - \cos{\varepsilon} ) ~.
\end{cases}
\label{eq:lightconediameter}
\end{equation}

Solution of the Equation (\ref{eq:lightconediameter}) with respect to parameter $\sin{\beta}$ gives

\begin{equation}
\sin{\beta} = \sqrt{\dfrac{1 - \cos{\gamma}}{1 - \cos{\varepsilon}}}~.
\label{eq:sinbeta}
\end{equation}

The angle $\varepsilon$ can be estimated by means of a pulse profile from observations of radio pulsars (see Figure~\ref{fig3} and Equation~(\ref{eq:varepsilon})). We suppose that a diameter $a$ of the emission cone at a distance $d$ covers the circle (dotted) around the larger cone formed by the magnetic axis $\mu$ rotating around the spin axis $\omega$ of the NS. The diameter of the cone $a$ can be expressed in terms of the width of the observed pulse profile since the start and end times of the passage of the base of the emission cone through the observer correspond to the beginning (rise) and end (fall) of the curve in the pulse profile. We use data of the $w_{10}$ parameter from the pulsar catalog, which is a pulse width at 10$\%$ of the peak of intensity \cite{Manchester...2016yCat....102034M} supposing that $w_{10}$ approximately corresponds to the ``travel'' time ($t$) of the emission cone passing through an observer
\begin{equation}
\varepsilon = \omega_{\rm s}\,t = \frac{2\pi }{P_{\rm s}} t \simeq \frac{2\pi }{P_{\rm s}} w_{10}~.
\label{eq:varepsilon}
\end{equation}

\vspace{-6pt}

\subsection{Data Selection and Evaluation of $B_{ns}(\beta)$}
\label{subsec:materils}

We use data from the ATNF pulsar catalog, available online\endnote{\url{https://www.atnf.csiro.au/people/pulsar/psrcat/} (accessed on 2 June 2023)} to apply our method to interested pulsars. The catalog contains information on rotation-powered pulsars and counts for over $3000$~objects, and it is the most extensive database that provides information about radio pulsars. The catalog is maintained by the Australia Telescope National Facility (ATNF). The catalog includes detailed information about pulsars, such as their positions, rotation periods, spin-down rates, dispersion measures, and other relevant parameters. The ATNF pulsar catalog was initially compiled using data from the Parkes radio telescope in Australia. Over time, data from other radio telescopes in Australia and worldwide were incorporated into the catalog. The catalog is regularly updated as new observations are made, and new pulsars are discovered; see \citep{Manchester...2005AJ...129...1993M} for more details.

The data selection was carried out according to the following criteria:
\begin{itemize}
\item Objects with known spin period $P_{\rm s}$; 
\item Exclusion of objects with $P_{\rm s} > 2\,\text{s}$;
\item Known spin-down rate $\dot{P}_{\rm s}$;
\item Known $w_{10}$ parameter.
\end{itemize}

The sample resulted in $1468$~objects from the ATNF pulsar catalog, which are NS with $P_{s}<2\,{\rm s}$ and the above parameters presented.

Some parameters of NS, such as radius $R_{\rm ns}$, mass $M_{\rm ns}$, and moment of inertia $I$ are in a narrow range close to canonical values; their variations should influence the evaluation insignificantly \cite{Lipunov...1992ans..book.....L, Lyne...2012puas.book.....L}. Therefore, in our calculations, we use canonical values of mass, radius, and moment of inertia as $R_{\rm ns} \simeq 10^{6}\,\text{cm}$, $M_{\rm ns} \simeq 1.4M_{\odot}$, $I \simeq 10^{45}\, \text{g cm}^2$, correspondingly.

\section{Results}
\label{sec:results}

Table~\ref{tab1} (in full available in a machine-readable format) summarizes our calculations of $\beta$ and $B_{\rm ns}(\beta)$ parameters for the chosen population. The statistics on calculated $\beta$ and $B_{\rm ns}(\beta)$ is given in Tables~\ref{tab2} and \ref{tab3}.

\begin{table}[H] 
\caption{Sample of radio pulsars with estimated $\beta$ and $B_{\rm ns}(\beta)$. Other parameters (coordinates, $P_{\rm s}$, $\dot{P}_{\rm s}$, $w_{10}$) are extracted from ATNF pulsar catalog~\cite{Manchester...2016yCat....102034M}. The full table containing data on 1468 objects is available online in a machine-readable format. \label{tab1}}
\begin{adjustwidth}{-\extralength}{0cm}
\begin{tabularx}{\fulllength}{llCCCCCCC}
\toprule
\textbf{No.}	& \textbf{Name}	& \textbf{RA} & \textbf{DEC} & \boldmath{$P_{\rm s}$} & \boldmath{$\dot{P}_{\rm s}$} & \boldmath{$w_{10}$} & \boldmath{$\beta$} &\boldmath{$B_{\rm ns}(\beta)$}\\
    & \textbf{PSR}  &\textbf{ J2000}  & \textbf{J2000} & \textbf{(s)} & \textbf{(s/s)} & \textbf{(ms)} & \textbf{(deg)} &\boldmath{($\times 10^{12}$ G)}\\
\midrule
1    &   J0006+1834 &  00:06:04.8 &   +18:34:59.0 &  0.69 &   2.10e-15 &   195.0 &     1.29 &  54.3 \\
2    &     B0011+47 &  00:14:17.7 &  +47:46:33.4 &  1.24 &  5.64e-16 &   142.5 &     2.11 &   23 \\
3    &   J0026+6320 &  00:26:50.5 &  +63:20:00.8 &  0.32 &  1.51e-16 &    48.0 &     3.22 &  3.94 \\
4    &     B0031$-$07 &  00:34:08.8 &  $-$07:21:53.4 &  0.94 &  4.08e-16 &   120.0 &      2.20 &  16.4\\
5    &   J0038$-$2501 &  00:38:10.2 &  $-$25:01:30.7 &  0.26 &   7.60e-19 &    15.0 &     9.01 &   0.093 \\
...  &         ...  &        ...  &         ...  &   ... &     ...   &   ...   &     ...   &    ... \\
\bottomrule
\end{tabularx}
\end{adjustwidth}
\end{table}

\vspace{-12pt}

\begin{table}[H] 
\caption{Statistics for the $\beta$-parameter, an angle between spin and magnetic dipole axes, for different groups of radio pulsars. \label{tab2}}
\begin{adjustwidth}{-\extralength}{0cm}
\begin{tabularx}{\fulllength}{CCCCCCC}
\toprule
\textbf{Puls. Group}	& \boldmath{$\beta$} \textbf{Min}	& \boldmath{$\beta$} \textbf{Max} & \boldmath{$\beta$} \textbf{Mean} & \boldmath{$\beta$} \textbf{Median} & \boldmath{$\sigma$} & \textbf{Total} \\
\midrule
$P_s < 0.01$ &  9.41 &   68.78 &  25.15  &   20.92  &   13.93 &   94 \\
$0.01 < P_s < 0.1$ & 3.22 &   61.84 &  13.33 &  9.46 &   10.51 &   73 \\
$0.1 < P_s < 2$ &  0.61 &  47.51 &  7.83 &  6.98 &    4.99 &  1301 \\
\midrule
All & 0.61 & 68.78 & 9.21 & 7.43 & 7.65 & 1468 \\
\bottomrule
\end{tabularx}
\end{adjustwidth}
\end{table}

\begin{table}[H] 
\caption{Statistics for the surface magnetic field strength $B_{\rm ns}(\beta)$ taking into account the $\beta$-parameter for different groups of radio pulsars. \label{tab3}}
\begin{adjustwidth}{-\extralength}{0cm}
\begin{tabularx}{\fulllength}{CCCCCCC}
\toprule
\textbf{Puls. Group}	& \boldmath{$B_{\rm ns}^{\rm min}(\beta)$}	& \boldmath{$B_{\rm ns}^{\rm max}(\beta)$} & \boldmath{$B_{\rm ns}^{\rm mean}(\beta)$} & \boldmath{$B_{\rm ns}^{\rm med}(\beta)$} & \boldmath{$\sigma$} & \textbf{Total} \\
\midrule
$P_s < 0.01$ &  7.76 $\times$ 10\textsuperscript{7} &  1.19 $\times$ 10\textsuperscript{10} &  1.92 $\times$ 10\textsuperscript{9}   &   5.75 $\times$ 10\textsuperscript{8}  &   1.68 $\times$ 10\textsuperscript{9} &   94 \\
$0.01< P_s < 0.1$ & 1.38 $\times$ 10\textsuperscript{9} 	 &   7.62 $\times$ 10\textsuperscript{13} &   	5.28 $\times$ 10\textsuperscript{12} &  2.96 $\times$ 10\textsuperscript{10} &   1.18 $\times$ 10\textsuperscript{13} &   73 \\
$0.1 < P_s < 2$ &  2.19 $\times$ 10\textsuperscript{10} &  7.56 $\times$ 10\textsuperscript{14} &  1.69 $\times$ 10\textsuperscript{13} & 8.33 $\times$ 10\textsuperscript{12}  & 3.39 $\times$ 10\textsuperscript{13} &  1301 \\
\midrule
All & 7.76 $\times$ 10\textsuperscript{7} & 7.56 $\times$ 10\textsuperscript{14} & 1.53 $\times$ 10\textsuperscript{13} & 7.1 $\times$ 10\textsuperscript{12} & 3.24 $\times$ 10\textsuperscript{13} & 1468 \\
\bottomrule
\end{tabularx}
\end{adjustwidth}
\end{table}

In the latter tables, we subdivided radio pulsars into three categories according to their spin periods to clarify and further underline the difference in evolutionary stages in Section~\ref{sec:discussion}.

For the population of middle-periodic pulsars ($0.1\text{s}<P_{\rm s}<2 \text{s}$) counting to \mbox{$1301$ known} objects, their values of $B_{\rm ns}(\beta)$ lie within the range $\sim$$10^{10}$--$10^{14}\, \text{G}$. Average values of $B_{\rm ns}(\beta)$ are in good agreement with the canonical value of magnetic field $10^{12}$--$10^{13}\, \text{G}$ for radio pulsars \cite{Igoshev...2021Univ....7..351I}. Their $\beta$ lie in the wide range $0.61$--$47.51\,\text{deg}$, but for most cases does not exceed $10$~deg, the median value of $\beta$ for the population of middle-periodic pulsars corresponds to 6.98 deg.

Short-periodic pulsars ($P_{\rm s} < 0.1 \text{s}$), including $94$ millisecond pulsars, altogether count to $167$~known objects. Their $B_{\rm ns}(\beta)$ values lie in the wide range $\sim$$10^{8}$--$10^{14}\,\text{G}$, but when considering millisecond pulsars only, their $B_{\rm ns}(\beta)$ cover $\sim$$10^{8}$--$10^{10}\,\text{G}$ range with \mbox{$5.8\times10^{8} \,\text{G}$} median value. Unlike the population of middle-periodic pulsars, millisecond objects have large values of $\beta$-parameter lying within the range $9.41$--$68.78\,\text{deg}$.

\begin{figure}[H]
\includegraphics[width=9.2 cm]{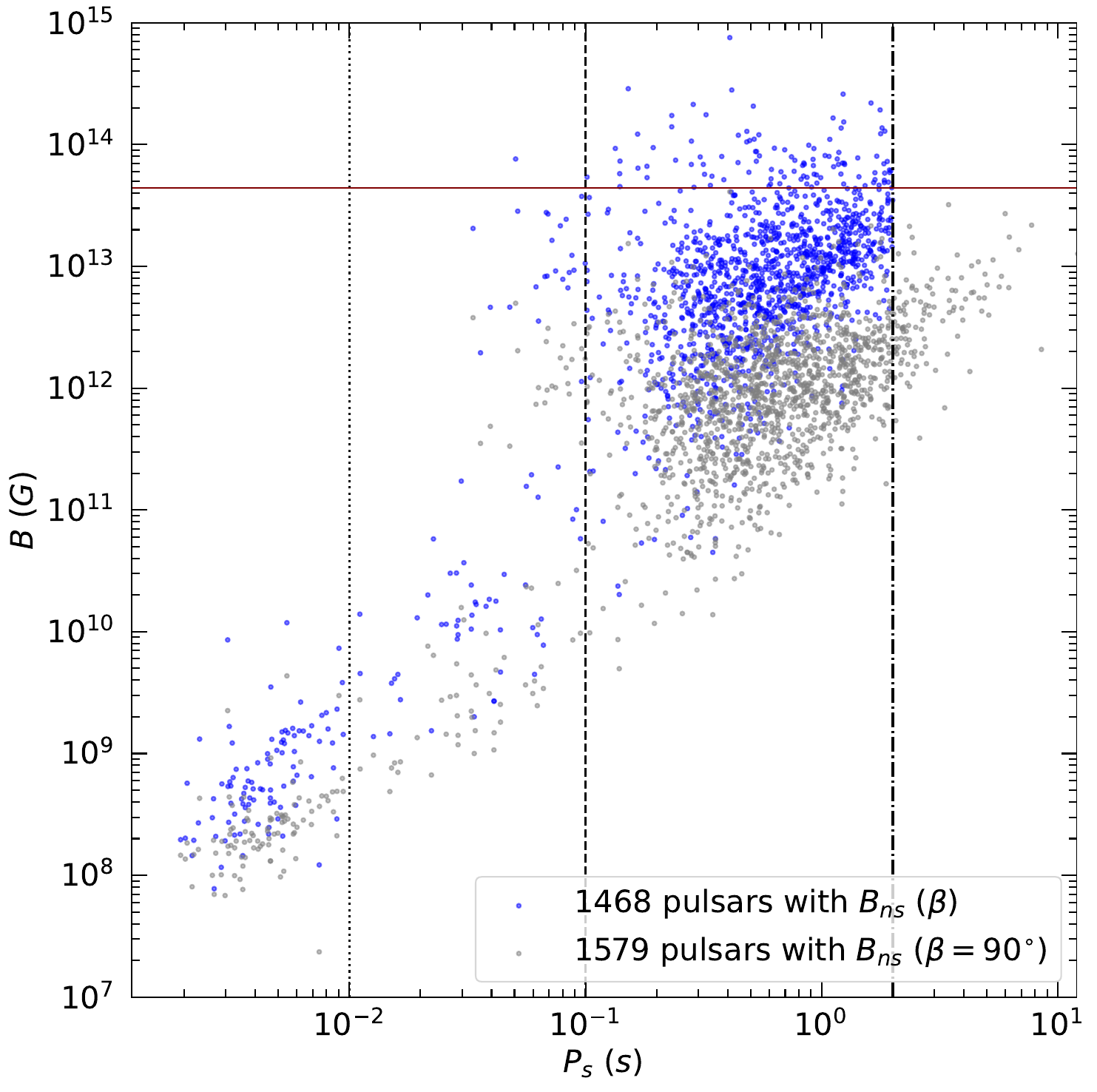}
\caption{Two data sets: $P_{\rm s}$--$B_{\rm ns}$ with (blue dots, calculated) and without (grey dots, retrieved from ATNF) taking into account $\beta$ parameter for radio pulsar population. The dotted vertical line corresponds to the borderline of $P_{\rm s} = 0.01\,\text{s}$ for the millisecond pulsar population. The dashed line corresponds to $P_{\rm s} = 0.1\,\text{s}$ borderline for the short-periodic pulsar population. The dash-dotted line corresponds to $P_{\rm s} = 2\,\text{s}$ borderline separating long-periodic pulsars. Blue dots are limited to $P_{\rm s} < 2\,\text{s}$ since we consider only rotation-powered pulsars with MDR mechanism of their energy loss (see Section~\ref{sec:introduction}). A horizontal solid red line corresponds to the quantum critical threshold $B_{\rm cr} \sim 4.4 \times 10^{13}\,\text{G}$ (see Section~\ref{sec:results}).}
\label{fig5}
\end{figure}

\begin{figure}[H]
\includegraphics[width=9.2 cm]{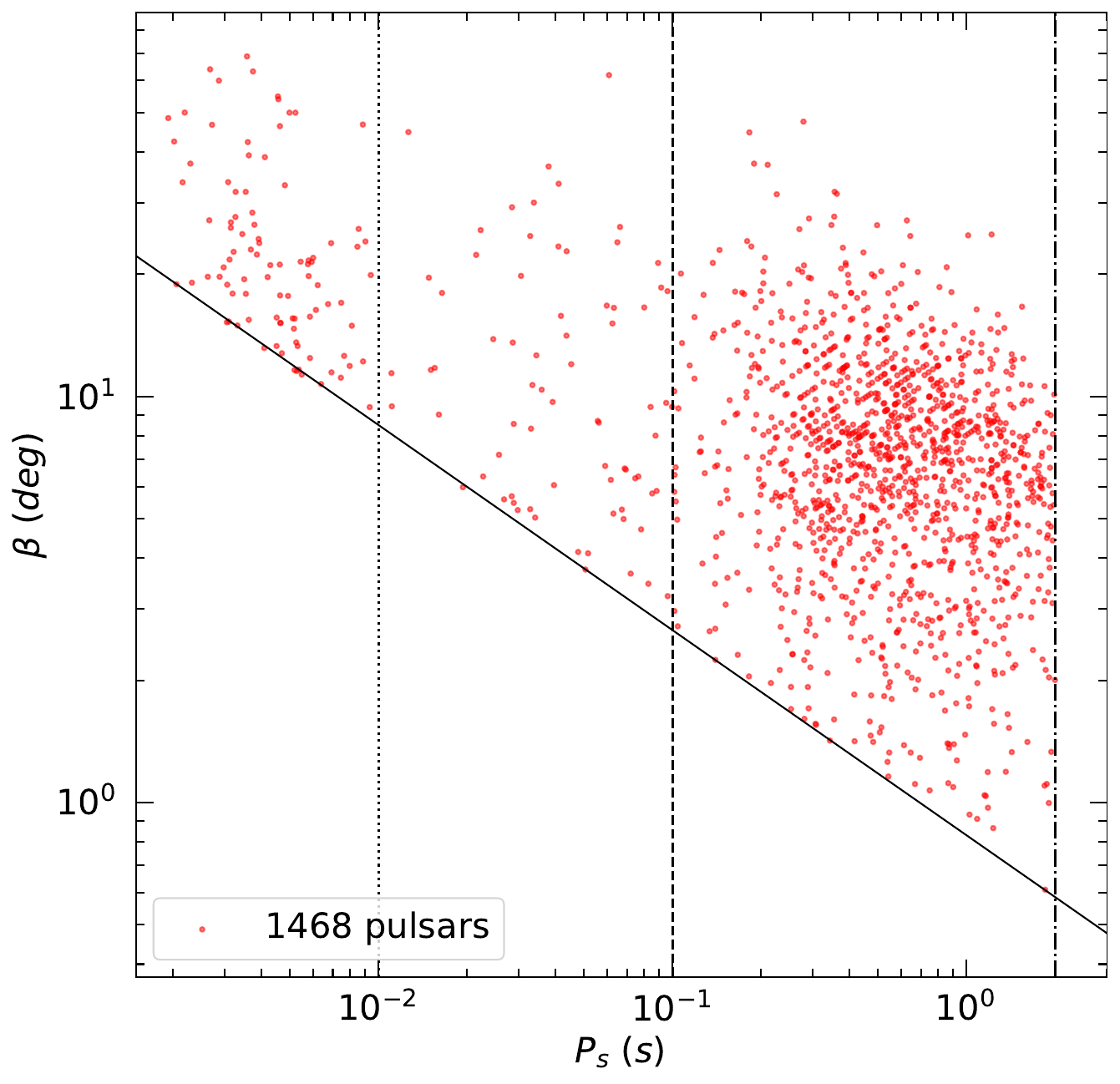}
\caption{Distribution of $\beta$-parameter depending on $P_{\rm s}$ for radio pulsar population with $P_{\rm s}<2~{\rm s}$. The solid black line corresponds to $\beta = f(P_{\rm s})$ with condition $w_{10} = P_{\rm s}/2$. Other lines are identical to those in Figure~\ref{fig5}. For all objects in the ATNF pulsar catalog, their observable $w_{10}$-parameters do not exceed half of the spin period, i.e., $w_{10} \leq P_{\rm s}/2$.}
\label{fig4}
\end{figure}

We built Figures~\ref{fig5} and \ref{fig4} on derived values of $B_{\rm ns}(\beta)$ and $\beta$ to show their general trend in relation to pulsars' spin periods. We distinguish the above three groups by vertical lines on both plots. In Figure~\ref{fig5}, we used two data sets as blue dots ($1468$~objects) for the calculated $B_{\rm ns}(\beta)$ and gray dots ($1579$~objects) for $B_{\rm ns}$ retrieved from the ATNF pulsar catalog with $\beta$ fixed at $90$~deg. Here, we sharply cut off data points with $P_{\rm s}>2$~s, thus eliminating long-periodic pulsars. As mentioned in Section~\ref{sec:introduction}, the primary mechanism of their rotational energy loss is the generation of pulsar wind
\cite{Malov...2022ARep...66...25M, Kenko...2023MNRAS.tmp.1064K}. In such a case, the $\dot{E}$ value does not depend on the $\beta$. Within this approach, the magnetic field of the NS can be estimated by knowing the power of the ejected pulsar wind, which cannot be estimated directly from observations. Therefore, the estimation of the magnetic fields of long-periodic radio pulsars ($P_{\rm s} > 2\text{s}$) is a model-dependent task and is beyond the scope of the current article.

Figure~\ref{fig4} shows the distribution of $1468$ derived $\beta$-parameters according to their spin periods. Two distinct features can be noticed: (a) some dots pull in a chain showing the positive trend sequences positioned parallel to each other, and (b) dots distribution generally goes above some level, here marked as a solid black line. Both features are related to the $w_{10}$-parameter, where the former points to the objects with similar values of $w_{10}$. At the same time, the latter peculiarity indicates that all pulsars in our sample obey the condition $w_{10} \leq P_{\rm s}/2$, i.e., each pulse duration does not exceed half of the spin period.

\vspace{-12pt}

\section{Discussion}
\label{sec:discussion}

In our paper, we used the classical dipole model of the radio pulsar magnetosphere proposed by \cite{Goldreich...1969ApJ...157..869G}. In this case, the magnetosphere of a neutron star has a dipole structure co-rotating with a pulsar. It is limited by the so-called light cylinder on which the linear velocity of the magnetic field lines reaches the speed of light. This model is canonical and relevant to this day \cite{Rahaman...2022MNRAS.516.3715R}.

Indeed, the width of the pulse profile of radio pulsars can vary depending on the frequency (wavelength) of the observed flux. However, significant deviations in the profile width are observed at lower frequencies ($<$200~MHz). According to \cite{Pilia...2016A&A...586A..92P}, this phenomenon is present because the light cone becomes wider when observed at lower frequencies, thereby seeing areas further from the pulsar's surface where the opening angle of the closed magnetic field lines is becoming broader. However, for higher frequencies ($>$200~MHz), this effect can be neglected \cite{Pilia...2016A&A...586A..92P}. The values of $w_{10}$ in the ATNF Pulsar Catalog are the average profile width in the range of frequencies between $400$--$2000$~MHz.

We showed that estimates of the surface magnetic field strength ($B_{\rm ns}$) for a population of known neutron stars in the radio pulsar (ejector) stage should depend essentially on the angle $\beta$ between spin and magnetic dipole axes of a neutron star. These estimates may differ by order of magnitude from those without considering the angle $\beta$ (see Figure~\ref{fig5}). The proposed method can be used when considering only rotation-powered pulsars with a MDR energy loss mechanism. This is not the case for the long-periodic pulsars with $P_s > 2$~s; therefore, we sharply cut off such objects in Figure~\ref{fig5}, although borderline transition cases may occur individually.

Within the framework of the proposed technique, it is not possible to estimate the evolution of $\beta$ over time since these changes are associated with changes in the flow of currents in the core of NS and the interaction of the magnetosphere with the surrounding plasma \citep{Philippov...2014MNRAS.441.1879P}. Nevertheless, we can compare our results against angles obtained within the framework of other methods.

As was mentioned in Section~\ref{sec:introduction} there are several approaches for estimating the $\beta$-parameter. They can be conditionally divided into two groups: geometric and polarimetric methods. The first is based on different geometric models for NS magnetic field and emission cone. Our method also belongs to the first group. The second is based on measuring the position angle of linear polarization from radio pulsars, which depends on $\beta$ \cite{Radhakrishnan...1969ApL.....3..225R}. Interest in comparing $\beta$ from these two approaches resulted in the following consideration.

\begin{figure}[H]
\includegraphics[width=0.95\textwidth]{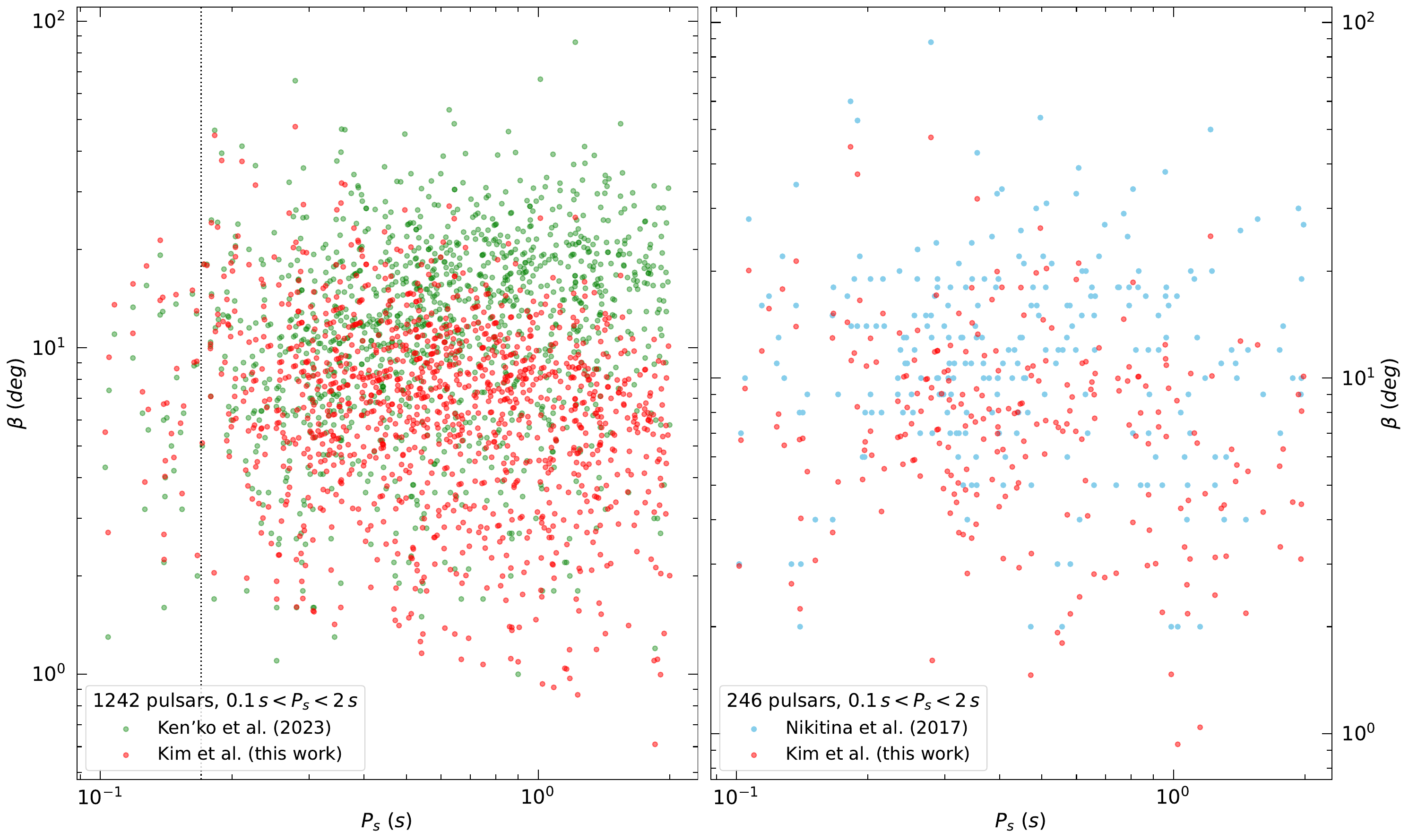}
\caption{A comparison of $\beta$-parameter estimation for a radio pulsar population obtained by Ken'ko~et~al.~(2023) \cite{Kenko...2023MNRAS.tmp.1064K} (green dots, $1242$~objects, \textbf{left} panel) and Nikitina et al. (2017) \cite{Nikitina...2017ARep...61..591N} (light blue dots, $246$~objects, \textbf{right} panel), and by our method (red dots, both panels). There are negative and positive trends between our data and data by Ken'ko et al. (2023) (\textbf{left} panel) correspondingly, with a vertical dotted line approximately marking the spin period where two methods give similar $\beta$ estimations. No trends are seen between our data and the data by Nikitina et al. (2017) (\textbf{right} panel).}
\label{fig4a}
\end{figure}

In recent articles \cite{Kenko...2022ARep...66..669K, Kenko...2023MNRAS.tmp.1064K, Nikitina...2017ARep...61..591N} an estimation of $\beta$-parameter was obtained within geometric method based on spherical trigonometry. A polar cap model was used by authors with an assumption that the line-of-sight passes through the center of the emission cone. The comparison between our data and data from \cite{Kenko...2022ARep...66..669K, Kenko...2023MNRAS.tmp.1064K, Nikitina...2017ARep...61..591N} is shown in Figure~\ref{fig4a} for matched $1242$~and~$246$~radio pulsars and their statistics are given in Table~\ref{tab:beta-comparison-kenko-nikitina-kim}. In most cases, the difference in estimates ($\Delta\beta$ median) does not exceed $5$~deg and is mainly caused by the difference in the methods (models) used. Negative and positive trends can be noticed correspondingly between our data and data by Ken'ko et al. (2023) \cite{Kenko...2023MNRAS.tmp.1064K} (Figure~\ref{fig4a}, left panel), with a vertical dotted line approximately marking the spin period where two methods give similar $\beta$ estimation. Again, this is due to differences in geometric approaches for $\beta$ estimation since the data themselves for both methods were taken from the same catalog (ATNF Pulsar Catalogue). No trends are seen between our data and the data by Nikitina et al.~(2017) \cite{Nikitina...2017ARep...61..591N} (Figure~\ref{fig4a}, right panel),  where blue dots are systematically positioned above red ones for matches pulsars. While these authors use analogous methods based on spherical trigonometry, their sample is relatively small, so the trends may not have enough data to manifest. Another reason could be in the data themselves, since in \cite{Nikitina...2017ARep...61..591N} the authors have used data from their observational facilities (Pushchino Radio Astronomy Observatory).

\vspace{-6pt}
\begin{table}[H] 
\caption{\textls[-15]{Statistics for the comparison of $\beta$ obtained from geometric approaches from Ken'ko~et~al.~(2023) \citep{Kenko...2023MNRAS.tmp.1064K} and Nikitina~et~al.~(2017) \citep{Nikitina...2017ARep...61..591N} with our data. All pulsars have spin periods between $0.1$~s and $2$~s}. \label{tab:beta-comparison-kenko-nikitina-kim}}
\begin{adjustwidth}{-\extralength}{0cm}
\begin{tabularx}{\fulllength}{cccccccc}
\toprule
\textbf{Pulsar Sample} & \textbf{Selection Criteria} & \textbf{Pulsar Number} & \boldmath{$|\Delta\beta|$} \textbf{Min}	& \boldmath{$|\Delta\beta|$} \textbf{ Max} & \boldmath{$|\Delta\beta|$} \textbf{ Mean} & \boldmath{$|\Delta\beta|$} \textbf{ Median} & \boldmath{$\sigma$} \\
\midrule
Ken'ko et al. (2023) \cite{Kenko...2023MNRAS.tmp.1064K} 
 & $P_s$, $w_{10}$ & $1242$ & $0.007338$ & $61.145636$ & $5.825957$ & $4.61572$ & $5.013604$ \\
Nikitina et al. (2017) \cite{Nikitina...2017ARep...61..591N} & $P_s$, $\dot{P}_{\rm s}$, $w_{10}$ & $246$ & $0.037164$ & $40.488991$ & $5.544564$ & $4.094941$ & $5.092834$ \\ 
\bottomrule
\end{tabularx}
\end{adjustwidth}
\end{table}

We further attempted to compare estimates obtained by polarimetric studies to determine the angle $\beta$ performed for only a small part of the radio pulsar population (see Table~\ref{tabx}). 
This method is based on measuring the position angle of linear polarization and is more reliable than geometric approaches. However, for some objects, when observed in different wavelength ranges (frequencies), it can give a significant scatter, especially for larger $\beta$. For example, as shown in \cite{Nikitina...2017ARep...61..591N} for PSR~B1055-52 (aka J1057-5226) $\beta$-parameter estimation at 10 cm wavelength gives $\beta_{10{\rm -cm}} = 15$~deg, but estimation at 20 cm wavelength gives $\beta_{20{\rm -cm}} = 24$~deg; for PSR~B1702-19 (aka J1705-1906) $\beta_{10{\rm -cm}} = 49$~deg,  $\beta_{20{\rm -cm}} = 70$~deg, etc. The larger scatter in $|\beta_{\rm pol}-\beta_{\rm geom}|$ between geometric and polarimetric methods is mainly due to the assumption that the line-of-sight passes through the center of the base of the emission cone. Thus, the geometric estimates are the lower limits for the measured angle $\beta$ \cite{Kenko...2023MNRAS.tmp.1064K}.

\textls[-25]{As also seen in Figure~\ref{fig5}, we obtained $110$~objects (7.5\% from 1468 pulsar sample) with estimated magnetic fields exceeding the so-called quantum critical threshold \mbox{$\sim$$4.4 \times 10^{13}\,\text{G}$ \cite{Peng...2007MNRAS.378..159P}}}. These are blue dots over the solid red line, and all (except one short-periodic source PSR~B0540-69) belong to the population of middle-periodic pulsars. The maximal value $B_{\rm ns}(\beta) \simeq 7.56\times10^{14}\,\text{G}$ refers to the pulsar PSR~J1119-6127. According to \cite{Gogus...2016ApJ...829L..25G} this radio pulsar demonstrates episodic SGR-like high-energy bursts reaching $2.8\times10^{39}\,\text{erg s}^{-1}$ within $15$--$150\,\text{keV}$ range. The  magnetic field of the NS derived from analysis of PSR~J1119-6127 during its burst activity corresponds to $B_{\rm ns}\sim 10^{14}$\,\text{G} \cite{Gogus...2016ApJ...829L..25G} that agrees with our estimate within an order of magnitude. The analysis of the rest of the high-B sub-sample can be interesting from the point of a possible relation between high-B radio pulsars and the population of isolated X-ray pulsars \cite{Benl...2017MNRAS.471.2553B}: anomalous X-ray pulsars (AXP), soft gamma-repeaters (SGR), etc.

\begin{table}[H] 
\caption{Comparison of the obtained data from this work ($\beta_{\rm geom}$) with data obtained by polarization method ($\beta_{\rm pol}$) from Nikitina~et~al.~(2017) \cite{Nikitina...2017ARep...61..591N}.  \label{tabx}}
\begin{adjustwidth}{-\extralength}{0cm}
\begin{tabularx}{\fulllength}{llCCC}
\toprule
\textbf{No.}	& \textbf{Name}	& \boldmath{$\beta_{geom}$} & \boldmath{$\beta_{pol}$} & \boldmath{$|\beta_{pol} - \beta_{geom}|$} \\
    & \textbf{PSR}  & \textbf{(deg)}  & \textbf{(deg)} & \textbf{(deg)}  \\
\midrule
1 & J0108-1431 & 3.93 & 11 &  7.07 \\
2 & B0656+14   & 5.18 & 17 & 11.82 \\
3 & J0905-5127 & 14.30 & 22 & 7.70\\
4 & J1015-5719 &  2.25 &  5 &  2.75 \\
5 & B1055-52   & 6.62 & 15 & 8.38 \\
6 & J1349-6130 & 9.05 & 64 &  54.95\\
7 & J1355-5925 &  6.49 &  10 &  3.51 \\
8 & B1509-58   & 3.07 & 10 & 6.93 \\
9 & B1702-19   & 10.41 & 49 & 38.59 \\
10 & J1702-4310 &  5.73 & 11 &  5.27 \\
11 & J1723-3659 &  17.96 &  28 &  10.04 \\
12 & B1800-21 & 2.64 & 12 & 9.36 \\
13 & B1822-14 & 5.07 & 8 & 2.93\\

\bottomrule
\end{tabularx}
\end{adjustwidth}
\end{table}

As seen in Figure~\ref{fig4} and Table~\ref{tab1}, with increasing spin period $P_{\rm s}$, there is a tendency of the angle $\beta$ to decrease. This agrees with the current view of the spin evolution of \mbox{NS \cite{Philippov...2014MNRAS.441.1879P}}: older neutron stars have lengthier spin periods and smaller values of $\beta$, excepting a millisecond pulsar population. According to \cite{Philippov...2014MNRAS.441.1879P} on the timescales $10^{6}$--$10^{7}\, \text{yr}$ in the ejector stage a NS should align its magnetic and spin axes, i.e., the angle $\beta$ tends to zero. 

For the population of millisecond pulsars (MSPs), the evolution of the $\beta$-parameter may differ significantly from other radio-pulsar populations. The millisecond pulsars are neutron stars in close binary systems or descendants of close binary systems in the case of isolated MSPs, with a low-mass companion, where accretion flow from a normal companion recycled a NS to ultra-short spin periods \cite{Lorimer...2008LRR....11....8L}. Thus, MSPs are old neutron stars whose rotational evolution has gone all possible stages (\textit{ejector} $\rightarrow$ \textit{propeller} $\rightarrow$ \textit{accretor}) and then came back to the \textit{ejector} stage through accretion recycling \cite{Lorimer...2008LRR....11....8L}. 

According to \cite{Urpin...1998MNRAS.295..907U},  the initial ejector stage for a neutron star in a binary system (with a normal star companion) lasts $10^{5}$--$10^{6}\,\text{yr}$, that is much shorter than in the case of an isolated NS and order of magnitude shorter than the timescale needed for aligning magnetic and spin axes of NS in ejector stage (see previous paragraph). Therefore, a NS in a binary system can move on to the following evolutionary stages (\textit{propeller} and \textit{accretor}) from the ejector stage with a $\beta$-parameter, which is significantly different from a zero value. Moreover, according to \cite{Biryukov...2021MNRAS.505.1775B}, the magnetic and spin axes of a neutron star in the stage of accretion tend to an orthogonal position, i.e., $\beta$-parameter increases to 90 deg on the timescale $\sim$$10^{5}\,\text{yr}$. The maximal possible lifetime of a NS on the accretor stage in a low-mass binary system is comparable to the lifetime of its normal companion, $\sim$$(0.1$--$10)\times10^{9}\, \text{yr}$ \cite{Kiziltan...2010ApJ...715..335K}. It exceeds the orthogonalization timescale by several orders of magnitude, sufficient to increase the $\beta$-parameter significantly. Thus, MSPs are old neutron stars that demonstrate large values of $\beta$-parameter compared to other types of radio pulsars in the ejector stage.

\vspace{6pt} 

\supplementary{Table 1: Sample of radio pulsars with estimated $\beta$ and $B_{\rm ns}(\beta)$. Other parameters (coordinates, $P_{\rm s}$, $\dot{P}_{\rm s}$, $w_{10}$) are extracted from ATNF pulsar catalog~\cite{Manchester...2016yCat....102034M}.
}

\authorcontributions{
Conceptualization, writing---review and editing, project administration, V.K. and Y.A.; methodology, investigation, writing---original draft preparation, supervision, V.K.; validation, data curation, visualization, A.U.; formal analysis, resources, funding acquisition, Y.A. All authors have read and agreed to the published version of the manuscript.}

\funding{
This research has been funded by the Science Committee of the Ministry of Education and Science of the Republic of Kazakhstan (Grant~No.~AP09258811).} 

\dataavailability{Table~\ref{tab1} is available in full online in a machine-readable format [file ``catalog\_table\_1.dat''].}

\acknowledgments{Authors acknowledge Remo Ruffini and the organizers of the conference ``Prof. Remo Ruffini Festschrift'' for the invitation to give a talk and the support provided during the online meeting. We thank the Editors and Referees for their comments and suggestions to improve the presentation of the results.}

\conflictsofinterest{
The authors declare no conflict of interest. The funders had no role in the design of the study; in the collection, analyses, or interpretation of data; in the writing of the manuscript; or in the decision to publish the results.}

\abbreviations{Abbreviations}{
The following abbreviations are used in this manuscript:\\

\noindent 
\begin{tabular}{@{}ll}
ATNF & Australia Telescope National Facility \\
AXP & Anomalous X-ray pulsar \\
deg & Degree (unit) \\
G & Gauss (unit) \\
MDR & Magnetic Dipole Radiation\\
ms & Millisecond (unit) \\
MSP & Millisecond pulsar \\
NS & Neutron star \\
SGR & Soft gamma-repeater \\
s & Second (unit) \\
s/s & Seconds per second (unit) \\
yr & Year (unit) \\
\end{tabular}
}

\begin{adjustwidth}{-\extralength}{0cm}
\printendnotes[custom] 

\reftitle{References}


\begin{thebibliography}{999}

\bibitem[{Lyne} and {Graham-Smith}(2012)]{Lyne...2012puas.book.....L}
{Lyne}, A.; {Graham-Smith}, F.
\newblock {\em {Pulsar Astronomy}}; {Cambridge University Press}: Cambridge, UK,  2012.

\bibitem[{Manchester} \em{et~al.}(2005){Manchester}, {Hobbs}, {Teoh}, and
  {Hobbs}]{Manchester...2005AJ...129...1993M}
{Manchester}, R.N.; {Hobbs}, G.B.; {Teoh}, A.; {Hobbs}, M.
\newblock {The Australia Telescope National Facility Pulsar Catalogue}.
\newblock {\em  Astron. J.} {\bf 2005}, {\em 129},~1993--2006.
\newblock {\url{https://doi.org/10.1086/428488}}.

\bibitem[{Tan} \em{et~al.}(2018){Tan}, {Bassa}, {Cooper}, {Dijkema},
  {Esposito}, {Hessels}, {Kondratiev}, {Kramer}, {Michilli}, {Sanidas},
  {Shimwell}, {Stappers}, {van Leeuwen}, {Cognard}, {Grie{\ss}meier},
  {Karastergiou}, {Keane}, {Sobey}, and
  {Weltevrede}]{Tan...2018ApJ...866...54T}
{Tan}, C.M.; {Bassa}, C.G.; {Cooper}, S.; {Dijkema}, T.J.; {Esposito}, P.;
  {Hessels}, J.W.T.; {Kondratiev}, V.I.; {Kramer}, M.; {Michilli}, D.;
  {Sanidas}, S.;  et~al.
\newblock {LOFAR Discovery of a 23.5 s Radio Pulsar}.
\newblock {\em  Astrophys. J.} {\bf 2018}, {\em 866},~54.
\newblock {\url{https://doi.org/10.3847/1538-4357/aade88}}.

\bibitem[{Malov} and {Marozava}(2022)]{Malov...2022ARep...66...25M}
{Malov}, I.F.; {Marozava}, H.P.
\newblock {On Braking Mechanisms in Radio Pulsars}.
\newblock {\em Astron. Rep.} {\bf 2022}, {\em 66},~25--31.
\newblock {\url{https://doi.org/10.1134/S106377292202007X}}.

\bibitem[{Ken'ko} and {Malov}(2023)]{Kenko...2023MNRAS.tmp.1064K}
{Ken'ko}, Z.V.; {Malov}, I.F.
\newblock {Evolution of the angles between magnetic moments and rotation axes
  in radio pulsars}.
\newblock {\em Mon. Not. R. Astron. Soc.} {\bf 2023}, \emph{522}, 1826--1842.
\newblock {\url{https://doi.org/10.1093/mnras/stad1099}}.

\bibitem[{Gunn} and {Ostriker}(1969)]{Gunn...1969Natur.221..454G}
{Gunn}, J.E.; {Ostriker}, J.P.
\newblock {Magnetic Dipole Radiation from Pulsars}.
\newblock {\em Nature} {\bf 1969}, {\em 221},~454--456.
\newblock {\url{https://doi.org/10.1038/221454a0}}.

\bibitem[{Pacini}(1970)]{Pacini...1970Natur.226..622P}
{Pacini}, F.
\newblock {The Nature of Pulsar Radiation}.
\newblock {\em Nature} {\bf 1970}, {\em 226},~622--624.
\newblock {\url{https://doi.org/10.1038/226622a0}}.

\bibitem[{Kuz'min} and {Dagkesamanskaya}(1983)]{Kuzmin...1983PAZh....9..149K}
{Kuz'min}, D.K.; {Dagkesamanskaya}, I.M.
\newblock {Evaluation of the angles of magnetic axis relative to rotation axis
  of the Pulsars.}
\newblock {\em Astron. Lett.} {\bf 1983}, {\em 9},~149--154.

\bibitem[{Lyne} and {Manchester}(1988)]{Lyne...1988MNRAS.234..477L}
{Lyne}, A.G.; {Manchester}, R.N.
\newblock {The shape of pulsar radio beams.} {\em Mon. Not. R. Astron. Soc.} {\bf 1988},
  {\em 234},~477--508. {\url{https://doi.org/10.1093/mnras/234.3.477}}.

\bibitem[{Malov}(1990)]{Malov...1990SvA....34..189M}
{Malov}, I.F.
\newblock {Angle Between the Magnetic Field and the Rotation Axis in Pulsars}.
\newblock {\em Sov. Astron.} {\bf 1990}, {\em 34},~189.

\bibitem[{Ken'ko} and {Malov}()]{Kenko...2022ARep...66..669K}
{Ken'ko}, Z.V.; {Malov}, I.F.
\newblock {Comparison of the Angles between the Magnetic Moment and Rotation
  Axis for Two Groups of Radio Pulsars}.
\newblock {\em Astron. Rep.} \textbf{2022}, {\em 66},~669--692.
\newblock {\url{https://doi.org/10.1134/S1063772922090050}}.

\bibitem[{Igoshev} \em{et~al.}(2021){Igoshev}, {Popov}, and
  {Hollerbach}]{Igoshev...2021Univ....7..351I}
{Igoshev}, A.P.; {Popov}, S.B.; {Hollerbach}, R.
\newblock {Evolution of Neutron Star Magnetic Fields}.
\newblock {\em Universe} {\bf 2021}, {\em 7},~351.
\newblock {\url{https://doi.org/10.3390/universe7090351}}.

\bibitem[{Xie} and {Zhang }(2014)]{Xie...2014AN....335..775X}
{Xie}, Y.; {Zhang}, S.N.
\newblock {Rotational behaviors and magnetic field evolution of radio pulsars}.
\newblock {\em Astron. Nachrichten} {\bf 2014}, {\em 335},~775,
\newblock {\url{https://doi.org/10.1002/asna.201412108}}.

\bibitem[{Portegies Zwart} and {Van den
  Heuvel}(1999)]{Portegies...1999NewA....4..355P}
{Portegies Zwart}, S.F.; {Van den Heuvel}, E.P.J.
\newblock {The origin of single radio pulsars}.
\newblock {\em New Astron.} {\bf 1999}, {\em 4},~355--363.
\newblock {\url{https://doi.org/10.1016/S1384-1076(99)00029-9}}.

\bibitem[{Lipunov}(1992)]{Lipunov...1992ans..book.....L}
{Lipunov}, V.M.
\newblock {\em {Astrophysics of Neutron Stars}}; {Springer: Berlin/Heidelberg, Germany; New York, NY, USA},  1992.

\bibitem[{Goldreich} and {Julian}(1969)]{Goldreich...1969ApJ...157..869G}
{Goldreich}, P.; {Julian}, W.H.
\newblock {Pulsar Electrodynamics}.
\newblock {\em  Astrophys. J.} {\bf 1969}, {\em 157},~869.
\newblock {\url{https://doi.org/10.1086/150119}}.

\bibitem[{Manchester} \em{et~al.}(2016){Manchester}, {Hobbs}, {Teoh}, and
  {Hobbs}]{Manchester...2016yCat....102034M}
{Manchester}, R.N.; {Hobbs}, G.B.; {Teoh}, A.; {Hobbs}, M.
\newblock {\emph{VizieR Online Data Catalog: ATNF Pulsar Catalogue (Manchester+,
  2005)}};
\newblock VizieR Online Data Catalog:
 {2016}; p. B/psr.


\bibitem[{Rahaman} \em{et~al.}(2022){Rahaman}, {Mitra}, {Melikidze}, and
  {Lakoba}]{Rahaman...2022MNRAS.516.3715R}
{Rahaman}, S.M.; {Mitra}, D.; {Melikidze}, G.I.; {Lakoba}, T.
\newblock {Pulsar radio emission mechanism---II. On the origin of relativistic
  Langmuir solitons in pulsar plasma}.
\newblock {\em Mon. Not. R. Astron. Soc.} {\bf 2022},
  {\em 516},~3715--3727.
\newblock {\url{https://doi.org/10.1093/mnras/stac2264}}.

\bibitem[{Pilia} \em{et~al.}(2016){Pilia}, {Hessels}, {Stappers}, {Kondratiev},
  {Kramer}, {van Leeuwen}, {Weltevrede}, {Lyne}, {Zagkouris}, {Hassall},
  {Bilous}, {Breton}, {Falcke}, {Grie{\ss}meier}, {Keane}, {Karastergiou},
  {Kuniyoshi}, {Noutsos}, {Os{\l}owski}, {Serylak}, {Sobey}, {ter Veen},
  {Alexov}, {Anderson}, {Asgekar}, {Avruch}, {Bell}, {Bentum}, {Bernardi},
  {B{\^\i}rzan}, {Bonafede}, {Breitling}, {Broderick}, {Br{\"u}ggen}, {Ciardi},
  {Corbel}, {de Geus}, {de Jong}, {Deller}, {Duscha}, {Eisl{\"o}ffel},
  {Fallows}, {Fender}, {Ferrari}, {Frieswijk}, {Garrett}, {Gunst}, {Hamaker},
  {Heald}, {Horneffer}, {Jonker}, {Juette}, {Kuper}, {Maat}, {Mann}, {Markoff},
  {McFadden}, {McKay-Bukowski}, {Miller-Jones}, {Nelles}, {Paas},
  {Pandey-Pommier}, {Pietka}, {Pizzo}, {Polatidis}, {Reich}, {R{\"o}ttgering},
  {Rowlinson}, {Schwarz}, {Smirnov}, {Steinmetz}, {Stewart}, {Swinbank},
  {Tagger}, {Tang}, {Tasse}, {Thoudam}, {Toribio}, {van der Horst},
  {Vermeulen}, {Vocks}, {van Weeren}, {Wijers}, {Wijnands}, {Wijnholds},
  {Wucknitz}, and {Zarka}]{Pilia...2016A&A...586A..92P}
{Pilia}, M.; {Hessels}, J.W.T.; {Stappers}, B.W.; {Kondratiev}, V.I.; {Kramer},
  M.; {van Leeuwen}, J.; {Weltevrede}, P.; {Lyne}, A.G.; {Zagkouris}, K.;
  {Hassall}, T.E.;  et~al.
\newblock {Wide-band, low-frequency pulse profiles of 100 radio pulsars with
  LOFAR}.
\newblock {\em Astron. Astrophys.} {\bf 2016}, {\em 586},~A92.
\newblock {\url{https://doi.org/10.1051/0004-6361/201425196}}.

\bibitem[{Philippov} \em{et~al.}(2014){Philippov}, {Tchekhovskoy}, and
  {Li}]{Philippov...2014MNRAS.441.1879P}
{Philippov}, A.; {Tchekhovskoy}, A.; {Li}, J.G.
\newblock {Time evolution of pulsar obliquity angle from 3D simulations of
  magnetospheres}.
\newblock {\em Mon. Not. R. Astron. Soc.} {\bf 2014},
  {\em 441},~1879--1887.  
\newblock {\url{https://doi.org/10.1093/mnras/stu591}}.

\bibitem[{Radhakrishnan} and
  {Cooke}(1969)]{Radhakrishnan...1969ApL.....3..225R}
{Radhakrishnan}, V.; {Cooke}, D.J.
\newblock {Magnetic Poles and the Polarization Structure of Pulsar Radiation}.
\newblock {\em Astrophys. Lett.} {\bf 1969}, {\em 3},~225.

\bibitem[{Nikitina} and {Malov}(2017)]{Nikitina...2017ARep...61..591N}
{Nikitina}, E.B.; {Malov}, I.F.
\newblock {Magnetic fields of radio pulsars}.
\newblock {\em Astron. Rep.} {\bf 2017}, {\em 61},~591--611.
\newblock {\url{https://doi.org/10.1134/S1063772917070058}}.

\bibitem[{Peng} and {Tong}(2007)]{Peng...2007MNRAS.378..159P}
{Peng}, Q.H.; {Tong}, H.
\newblock {The physics of strong magnetic fields in neutron stars}.
\newblock {\em Mon. Not. R. Astron. Soc.} {\bf 2007},
  {\em 378},~159--162.
\newblock {\url{https://doi.org/10.1111/j.1365-2966.2007.11772.x}}.

\bibitem[{G{\"o}{\u{g}}{\"u}{\c{s}}}
  \em{et~al.}(2016){G{\"o}{\u{g}}{\"u}{\c{s}}}, {Lin}, {Kaneko}, {Kouveliotou},
  {Watts}, {Chakraborty}, {Alpar}, {Huppenkothen}, {Roberts}, {Younes}, and
  {van der Horst}]{Gogus...2016ApJ...829L..25G}
{G{\"o}{\u{g}}{\"u}{\c{s}}}, E.; {Lin}, L.; {Kaneko}, Y.; {Kouveliotou}, C.;
  {Watts}, A.L.; {Chakraborty}, M.; {Alpar}, M.A.; {Huppenkothen}, D.;
  {Roberts}, O.J.; {Younes}, G.;  et~al.
\newblock {Magnetar-like X-Ray Bursts from a Rotation-powered Pulsar, PSR
  J1119-6127}.
\newblock {\em  Astrophys. J. Lett.} {\bf 2016}, {\em 829},~L25.
\newblock {\url{https://doi.org/10.3847/2041-8205/829/2/L25}}.

\bibitem[{Benli} and {Ertan}(2017)]{Benl...2017MNRAS.471.2553B}
{Benli}, O.; {Ertan}, {\"U}.
\newblock {On the evolution of high-B radio pulsars with measured braking
  indices}.
\newblock {\em Mon. Not. R. Astron. Soc.} {\bf 2017},
  {\em 471},~2553--2557.
\newblock {\url{https://doi.org/10.1093/mnras/stx1735}}.

\bibitem[{Lorimer}(2008)]{Lorimer...2008LRR....11....8L}
{Lorimer}, D.R.
\newblock {Binary and Millisecond Pulsars}.
\newblock {\em Living Rev. Relativ.} {\bf 2008}, {\em 11},~8.
\newblock {\url{https://doi.org/10.12942/lrr-2008-8}}.

\bibitem[{Urpin} \em{et~al.}(1998){Urpin}, {Geppert}, and
  {Konenkov}]{Urpin...1998MNRAS.295..907U}
{Urpin}, V.; {Geppert}, U.; {Konenkov}, D.
\newblock {Magnetic and spin evolution of neutron stars in close binaries}.
\newblock {\em Mon. Not. R. Astron. Soc.} {\bf 1998},
  {\em 295},~907--920.
\newblock {\url{https://doi.org/10.1046/j.1365-8711.1998.01375.x}}.

\bibitem[{Biryukov} and {Abolmasov}(2021)]{Biryukov...2021MNRAS.505.1775B}
{Biryukov}, A.; {Abolmasov}, P.
\newblock {Magnetic angle evolution in accreting neutron stars}.
\newblock {\em Mon. Not. R. Astron. Soc.} {\bf 2021},
  {\em 505},~1775--1786.
\newblock {\url{https://doi.org/10.1093/mnras/stab1378}}.

\bibitem[{Kiziltan} and {Thorsett}(2010)]{Kiziltan...2010ApJ...715..335K}
{Kiziltan}, B.; {Thorsett}, S.E.
\newblock {Millisecond Pulsar Ages: Implications of Binary Evolution and a
  Maximum Spin Limit}.
\newblock {\em  Astrophys. J.} {\bf 2010}, {\em 715},~335--341.
\newblock {\url{https://doi.org/10.1088/0004-637X/715/1/335}}.

\end{thebibliography}

\PublishersNote{}
\end{adjustwidth}
\end{document}